\documentclass[12pt]{iopart}

\usepackage{epsfig}

\begin{document}

\title{Mean-field approximations for the restricted solid-on-solid 
growth models}

\author{Andre C. Barato and M\'{a}rio J. de Oliveira}

\address{Instituto de F\'{\i}sica,
Universidade de S\~{a}o Paulo, \\
Caixa Postal 66318\\
05315-970 S\~{a}o Paulo, S\~{a}o Paulo, Brazil}

\ead{oliveira@if.usp.br}

\begin{abstract}

We study models for surface growth with a wetting and a roughening 
transition using simple and pair mean-field approximations.
The simple mean-field equations are solved exactly and they
predict the roughening transition and the correct growth exponents
in a region of the phase diagram. The pair mean-field equations, 
which are solved numerically, show a better
accordance with numerical simulation and correctly predicts a growing
interface with constant velocity at the moving phase. 
Also, when detailed balance is fulfilled, the pair mean 
field becomes the exact solution of the model.

\end{abstract}


\maketitle

\section{Introduction}

Surface growing \cite{bara95,krug97}
is a phenomenon observed in nature as
well as in the laboratory. As an example of the latter
we cite the experimental technique known as molecular 
beam epitaxy which allows us the growing of surface
at the atomic level. 
The surface is
described by a height variable $h({\bf r},t)$ that gives the
height of the deposited layer at a given point ${\bf r}$ of
the substrate at time $t$. 
Several models have
been introduced to describe the mean features of
surface growing in which the heights 
are stochastic variables 
whose time evolution is governed by a Markovian stochastic process.
Here we are concerned with the solution of such models
by the use of mean-field approximations.

Two relevant quantities are used to characterize the
surface growth. One is the mean height ${\bar h}$ of the surface from 
the substrate and the other is the surface width $w$,
which is a measure of the surface roughness.
The divergence of $w$ characterizes a rough interface. 
According to Family and Vicsek \cite{fami85}, the width of a 
rough surface of a sufficient large system behaves as
\begin{equation}
w \sim t^{\,\gamma},
\label{1}
\end{equation}
where $\gamma$ is the growth exponent.
This behavior characterizes a rough thermodynamic phase. Otherwise, 
that is, if the width of an infinite system remains finite when
$t\to\infty$, the surface is smooth.
A roughening transition takes place when, 
by varying the control parameters, the surface 
changes from a smooth to a rough surface.

The interface may yet be pinned 
or moving. A moving thermodynamic phase is characterized
by a constant velocity $v$ of the interface, or in other
words, by a linear growth of the mean height, that is,
\begin{equation}
{\bar h}= v t.
\label{2}
\end{equation}
A depinning transition \cite{reis03} from a 
pinned to a moving phase occurs when, by changing the control
parameters, the interface begins to move with a constant velocity.
This is also known as a wetting transition \cite{diet86} since 
in the moving phase the mean height becomes infinitely large when 
$t\to\infty$, that characterizes a wet phase.
At the depinning transition, the mean height
may not grow linearly with time but may behave according to
\begin{equation}
{\bar h} \sim t{\,^\gamma}.
\label{3}
\end{equation}
In the model we study here the depinning transition 
coincide with the roughening transition. 

If we define $P(h,t)$ as the one-point height probability distribution
at time $t$, the mean height ${\bar h}$ and the square $w^2$ of the 
surface width are the first moment and the variance 
of this probability distribution, respectively. The asymptotic behavior
given by Eqs. (\ref{1}) and (\ref{2}) can then be obtained from  
the following scaling form \cite{gine04}
\begin{equation}
P(h,t) = t^{-\gamma} f(\frac{h-vt}{t^{\gamma}}),
\label{scal}
\end{equation}
where $f(x)$ is a universal function.
The behaviors given by Eqs. (\ref{1}) and (\ref{3})  
also follows from this same form by setting $v=0$.

Some models have been introduced to describe the
surface growth and the asymptotic behavior in which the heights
are stochastic variables governed by Langevin equations.
Two important ones are the Edwards-Wilkinson (EW) \cite{edwa82}
\begin{equation}
\frac{\partial h}{\partial t}= \nu \nabla^2h+ 
\eta({\bf r}, t),
\end{equation}
and the Kardar-Parisi-Zhang (KPZ) \cite{kard86} 
\begin{equation}
\frac{\partial h}{\partial t}= \nu \nabla^2h+ 
\lambda(\nabla h)^2+ \eta({\bf r},t),
\end{equation}
where $\eta({\bf r},t)$ is a white noise. 
Since the EW equation is linear, it stays invariant under the 
transformation $h\to-h$ while 
the KPZ equation, which is nonlinear, lacks this property.
In one dimension the growth exponent for 
the EW class is $\gamma= 1/4$ and for the KPZ class is $\gamma=1/3$.    

In this work we study a growth model, introduced by 
Hinrichsen et al. \cite{hinr97},
with deposition and
evaporation of particles that respects the restricted solid on
solid (RSOS) condition \cite{kost89} and in which 
the evaporation at the initial height is forbidden,
that is, there is a wall at the zero height.
The presence of the wall leads to a nonequilibrium wetting transition
depending on the evaporation and deposition rates.
It is worth mentioning that 
nonequilibrium wetting transition has been already studied
by another approach namely 
by mapping the KPZ with a potential into a Langevin
equation with a multiplicative noise \cite{tu97,muno98}.

The growth model is studied here 
by means of simple and pair mean-field approximations.
The mean-field approximation at the pair level,
which does respect the RSOS condition exactly, is found
to be capable of describing a moving phase, that is,
a moving interface at constant velocity.
Another feature of the two-site mean-field approximation
is that it becomes the exact solution when detailed balance
is fulfilled.

Recently, a mean-field theory for surface growth
has been introduced by Hinrichsen et al. \cite{hinr03}
and used by Ginelli and Hinrichsen \cite{gine04} to
study a single step model for surface growth. 
The one-site and two-site mean-field approximations we use here 
are distinct from that of Hinrichsen et al. \cite{hinr03}
but share properties that are similar.

The paper is organized as follows. In the next section 
we define the model to be examined and write down 
the master equation for general models with the RSOS condition.
In section III we solve the master equation within the 
one-site mean-field approach exactly. The section IV is
dedicated to the pair mean-field approach. The 
resultant equations are solved numerically and compared with the 
results obtained from the one-site approximation and from simulations. 
In section V we analyze the detailed balance condition and
in section VI we present our conclusions.    

\section{Model and master equation}

\begin{figure}
\centering
\epsfig{file=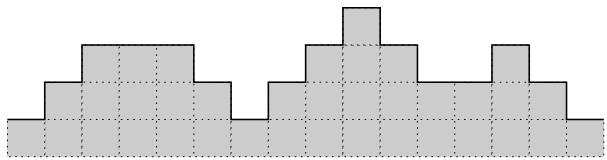,width=8cm}
\caption{A configuration of heights with the RSOS condition}
\label{rsos}	
\end{figure}

We consider a one-dimensional lattice where
a discrete variable $h_i$ is attached to the site $i$,
that represents the height of the pile of particles at site $i$.
The stochastic variable $h_i$ takes the integer
values. At time $t=0$ we consider a flat surface
at the zero level, that is,  $h_i=0$ for all $i$.
We consider models with random sequential updates
of the heights, that have not
only deposition but also evaporation.
We will restrict ourselves only to models
that obey the RSOS condition, that is, such that
$|h_i-h_{i+1}|\leq 1$ (see Fig. \ref{rsos}).
At each time step only one
site is updated according to local stochastic rules.
If site $i$ is chosen, these rules will affect only
the neighboring sites $i+1$ and $i-1$ and site $i$ itself. 
The rate of the transition $h_i\to h_i+n$ is denoted
by $c_n(h_{i-1},h_i,h_{i+1})$. 
The only possible transitions are those for which $n=\pm1$
meaning that the height is increased (deposition)
or decreased (evaporation) by just one unit.

\begin{figure}
\centering
\epsfig{file=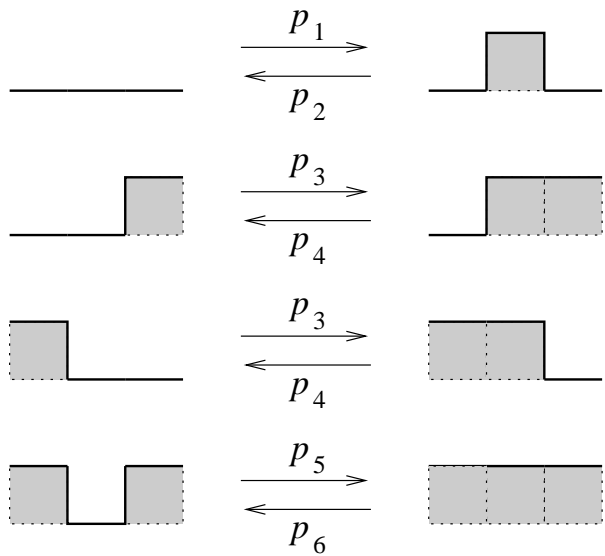,width=8cm}
\caption{Transition rates.}
\label{rate}	
\end{figure}

The possible transition rates are presented in Fig. \ref{rate}
and are given by
\begin{equation}
c_+(h,h,h) = p_1,
\end{equation}
\begin{equation}
c_+(h,h,h+1) = c_+(h+1,h,h) = p_3,
\end{equation}
\begin{equation}
c_+(h+1,h,h+1) = p_5,
\end{equation}
\begin{equation}
c_-(h-1,h,h-1) = p_2,
\end{equation}
\begin{equation}
c_-(h-1,h,h) = c_-(h,h,h-1) = p_4,
\end{equation}
\begin{equation}
c_-(h,h,h) = p_6,
\end{equation}
and the model so defined has six parameters: $p_1, \ldots, p_6$. 
By rescaling time we see that they are not all independent
and one of them can set equal to unity.

The probability $P(h_1,h_2,\ldots,t)$ 
of a given configuration $(h_1,h_2,\ldots)$ 
at time $t$ obeys the master equation
\[
\frac{d}{dt} P(h_1,h_2,\ldots,t) =
\]
\[
\sum_{n=\pm1}\sum_i 
\{ c_n(h_{i-1},h_i-n,h_{i+1}) P(h_1,\ldots,h_i-n,\ldots,t)  
\]
\begin{equation}
-c_n(h_{i-1},h_i,h_{i+1}) P(h_1,\ldots,h_i,\ldots,t) \}, 
\label{master}
\end{equation}
and we are considering periodic boundary conditions.
For late use we write down the time evolution
of the marginal probability distribution related 
to a given site, say site $1$. It is given by
\[
\frac{d}{dt} P(h_1,t) = \sum_{h_0,h_2}
\sum_{n=\pm1}
\{ c_n(h_0,h_1-n,h_2) P(h_0,h_1-n,h_2,t)  
\]
\begin{equation}
-c_n(h_0,h_1,h_2) P(h_0,h_1,h_2,t) \}.
\label{one} 
\end{equation}
We also write down the time evolution of the 
marginal probability distribution related to
two consecutive sites, say site 1 and 2. It is 
given by
\[
\frac{d}{dt} P(h_1,h_2,t) =
2 \sum_{h_0} \sum_{n=\pm1}
\{ c_n(h_0,h_1-n,h_2) P(h_0,h_1-n,h_2,t)
\]
\begin{equation}
-c_n(h_0,h_1,h_2) P(h_0,h_1,h_2,t) \},
\label{two}
\end{equation}
valid for solutions that are translationally invariant,
where we have made use of the symmetry 
$c_n(h_0,h_1,h_2)=c_n(h_2,h_1,h_0)$.

For certain values of the parameters the stochastic process
exhibits detailed balance. This means to say that each term
of the summation on the right hand side of Eq. (\ref{master})
vanishes in the time independent stationary state.
In this case the time independent stationary probability 
distribution can be written as the product 
\begin{equation}
P(h_1,h_2,h_3,h_4,\ldots) = \frac1Z\, T(h_1,h_2)\,T(h_2,h_3)
T(h_3,h_4)\ldots
\label{pt}
\end{equation}
where $T(h_1,h_2)$ are the elements of a symmetric matrix $T$,
that vanish whenever the RSOS condition is not fulfilled.
The detailed balance condition gives the following relations
for the nonvanishing elements of $T$, denoted by
$X_h = T(h,h)$ and $Y_h = T(h,h+1)$,
\begin{equation}
p_1\, X_h^2  = p_2\, Y_h^2,
\label{db1}
\end{equation}
\begin{equation}
p_3 \, X_h Y_h= p_4\, Y_h X_{h+1},
\label{db2}
\end{equation}
\begin{equation}
p_5\, Y_h^2 = p_6\, X_{h+1}^2.
\label{db3}
\end{equation}
From these equations follows the relation \cite{neer}
\begin{equation}
\frac{p_1\,p_5}{p_2\,p_6} = \frac{p_3^{\,2}}{p_4^{\,2}},
\label{db}
\end{equation}
which is the condition for detailed balance to hold.

In the specific model \cite{hinr97} we study here, 
the deposition occurs with a rate $q$ and the evaporation, for 
nonzero heights, occurs with rates $r$ or $p$, depending on the 
neighbors configuration. More specifically, 
if $h_i= 0$, $p_1=p_3=p_5=q$ and $p_2=p_4=p_6=0$; and
if $h_i\neq 0$, 
\begin{equation}
p_1=p_3=p_5=q, \qquad\qquad
p_2=p_4= r,    \qquad\qquad
p_6= p.
\label{sm}
\end{equation}
The detailed balance condition (\ref{db}) 
gives $p=r$. Therefore, when $p=r$
is respected the model can be solved exactly 
\cite{hinr97} for the pinned phase.

Without the wall at the initial height, we would have a rough 
interface growing in the positive direction for $q>q_c$ and in 
the negative direction for $q<q_c$. 
With the wall the phase $q>q_c$ 
is not affected, in the the sense that the interface still grows 
and is rough. For $q<q_c$, in the presence of the wall,
the interface is smooth and stays pinned to the substrate. 
Hence, at $q=q_c$, the model displays a depinning
and a roughening transition.
The rough interface may display a EW or a KPZ 
behavior. In the case $p=r$, it is known that the crossover from
EW to KPZ behavior \cite{reis06} coincides with the critical point
occurring at $q=p$.

The case $p=0$ \cite{alon96,alon98}
is special since in this case no particle can
be evaporated from a completed filled layer and 
the presence of the wall thus makes no difference. 
The critical behavior at $p=0$ places the model in the universality
class of the directed percolation (DP) class 
\cite{hinr00,marro99}.

The order parameter of the pinned phase is the density of sites,
$P_0$, in contact with the substrate, or the wall. 
It is assumed to behave near and below the transition as
\begin{equation}
P_0 \sim (q_c-q)^\beta.
\label{op}
\end{equation}
The interface width $w$ is finite at the pinned phase but 
diverges as one approaches the critical point. We
assume that it diverges according to 
\begin{equation}
w \sim (q_c-q)^{-\zeta}.
\end{equation}
The order parameter of the moving phase is chosen to be
the velocity $v$ of the interface growing, defined by
\begin{equation}
v=\frac{d}{dt}\langle h_1 \rangle = 
\sum_{h_1} h_1 \frac{d}{dt} P(h_1,t).
\end{equation}
which can also be written as 
\begin{equation}
v= \sum_{h_0,h_1,h_2}
\sum_{n=\pm1}
n c_n(h_0,h_1,h_2) P(h_0,h_1,h_2,t).
\end{equation}
Near and above the transition we assume that the velocity behaves as
\begin{equation}
v \sim (q-q_c)^\theta.
\end{equation}

\section{Simple mean-field approximation}

We are interested in studying the solution of the master
equation associated to an infinite system.
In order to solve the master equation (\ref{master}) 
we begin by using a simple mean
field approximation. In this approach all the correlations are
neglected resulting into the following approximation:
$P(h_{i-1}, h_i, h_{i+1},t)= P(h_{i-1},t)P(h_i,t)P(h_{i+1},t)$.
Insertion of this approximation into equation (\ref{one}) 
yields a closed equation for the one-site probability $P_k(t)$ 
distribution
\[
\frac{d}{dt}P_k= q(P_{k-1}^3- P_{k}^3)
+(r-2q)(P_k^2P_{k+1}- P_{k-1}^2P_k)
\]
\begin{equation}
+(2r-q)(P_kP_{k+1}^2-P_{k-1}P_k^2) + p(P_{k+1}^3- \gamma_k P_{k}^3),
\end{equation}
valid for the specific model defined by the rates (\ref{sm}),
where we are using the integer variable $k$ in the place of the height
$h$ and initially $P_0=1$ and $P_k=0$ for $k\neq0$
because we are considering a initial flat surface.
The variable $\gamma_k=0$ when $k=0$ and $\gamma_k=1$ otherwise.
The equation is valid for $k>0$ and for $k=0$ provided
we set $P_{-1}=0$ on the right-hand side.

\begin{figure}
\centering
\epsfig{file=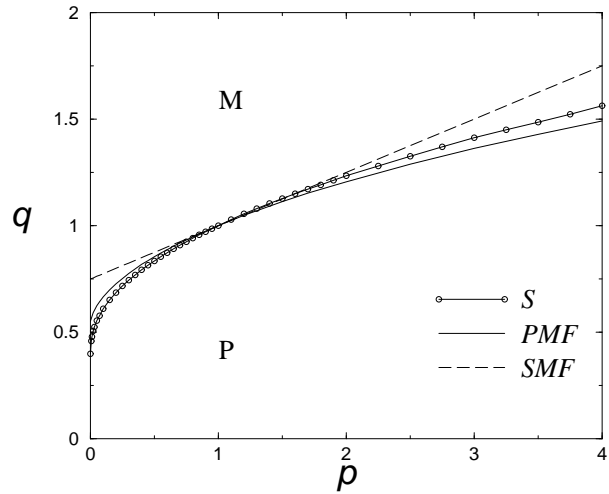,width=8cm}
\caption{Phase diagram for $r=1$ in the $p-q$ plane. 
The critical lines, separating the moving (M) and pinned (P) phases, 
were determined by simple mean-field (SMF), pair mean-field (PMF)
approximation and numerical simulations (S).}
\label{phase}	
\end{figure}

Let us first look for a possible stationary state, that is, 
a time-independent solution, that correspond to a pinned phase.
If we assume a solution of the form 
\begin{equation}
P_k=A\lambda^k,
\label{sol}
\end{equation}
we can check easily by substitution that this is indeed a solution
provided $\lambda$ is the root of the third-order algebraic equation
\begin{equation}
-p\lambda^3+(q-2r)\lambda^2+(2q-r)\lambda+q=0.
\end{equation}
The critical line, shown in Fig. \ref{phase}, 
is found by letting $\lambda\to 1$ with the result
\begin{equation} 
q_c=\frac14(p+3r),
\end{equation}
and the stationary solution occurs for $q<q_c$. 
The normalization of $P_k$, given by (\ref{sol}),
gives $A=1-\lambda$ so that
$P_0=1-\lambda$. The mean height ${\bar h}$ and the square of the
interface width $w^2$ are determined as the average 
and the variance 
of the distribution given by (\ref{sol}), that is
\begin{equation}
{\bar h} = \langle k \rangle
\qquad{\rm and}\qquad
w^2 = \langle k^2\rangle -{\bar h}^2,
\label{observ}
\end{equation}
resulting in
${\bar h}=\lambda/(1-\lambda)$
and $w=\sqrt{\lambda}/(1-\lambda)$. 
Near the critical line, $\lambda$ approaches 1 as
$1-\lambda = 2(q_c-q)/(r+p)$ from which follows the results
$P_0\sim (q_c-q)$, ${\bar h} \sim (q_c-q)^{-1}$ and
$w\sim (q_c-q)^{-1}$. 

We solve the equation for $q\geq q_c$ 
by a method similar to that of
Ginelli and Hinrichsen \cite{gine04} in which
a continuous version of the equation is used.
Writing $h=k$, the equation for the probability distribution
$P(h,t)$ becomes, up to second order,
\begin{equation}
\frac{\partial P}{\partial t} =
-a P^2 \frac{\partial P}{\partial h} 
+ b \left[ P\left(\frac{\partial P}{\partial h}\right)^2
+ \frac12 P^2\frac{\partial^2 P}{\partial h^2} \right],
\end{equation}
where 
\begin{equation}
a = 12 (q- q_c),
\end{equation}
and
\begin{equation}
b = 2q+r+3p.
\end{equation}

At the critical line, $q=q_c$, the coefficient $a$ vanishes
and we end up with the equation
\begin{equation}
\frac{\partial P}{\partial t} =
b \left[ P\left(\frac{\partial P}{\partial h}\right)^2 +
\frac12 P^2\frac{\partial^2 P}{\partial h^2} \right],
\end{equation}
which can be solved by assuming the scaling form 
(\ref{scal}) for $P(h,t)$. 
A consistency is achieved only
if we choose the growth exponent as being $\gamma=1/4$. 
Along the critical line, the simple mean-field approximation
is then compatible with an EW behavior.
The substitution of the scaling form
into the equation leads to the following equation for 
the scaling function $f(x)$
\begin{equation}
f(x) + xf^\prime(x) + 4bf(x) [f^\prime(x)]^2 
+ 2b[f(x)]^2 f^{\prime\prime}(x) = 0,
\end{equation}
whose solution is 
\begin{equation}
f(x) = \sqrt{1-\frac{x^2}{2b}}.
\end{equation}
The asymptotic probability distribution, valid for large times,
can then be written as
\begin{equation}
P(h,t) = \sqrt{\frac{1}{t^{1/2}}-\frac{h^2}{2bt}}.
\end{equation}

For $q>q_c$ it suffices to consider only the linear term in $b$.
The equation for $P(h,t)$ reads
\begin{equation}
\frac{\partial P}{\partial t} =
- a P^2 \frac{\partial P}{\partial h},
\end{equation}
where now $a$ is strictly positive.
This equation can be solved by assuming again a scaling relation
of the form (\ref{scal}). Now, however, the consistency requires
an exponent $\gamma=1/3$ so that the simple mean-field
approximation is then compatible with a KPZ behavior.
The substitution gives the following equation for the scaling
function
\begin{equation}
f(x)+xf^\prime(x) = 3a[f(x)]^2 f^\prime(x),
\end{equation}
whose solution is 
\begin{equation}
f(x) = \sqrt{\frac{x}{a}}.
\end{equation}

The asymptotic probability distribution,
valid for large time, can then be written as
\begin{equation}
P(h,t) = \sqrt{\frac{h}{at}}.
\end{equation}
We remark that $P(h,t)$ vanishes for values of $h$ 
larger than a certain $h_{\rm max}$ which depends on time.
This maximum value of $h$ is determined by the normalization
of $P(h,t)$ and is given by $h_{\rm max}=(9at/4)^{1/3}$.
This allows us to determine
${\bar h} = 3 h_{\rm max}/5$ and 
$w = (12/175)^{1/2}h_{\rm max}$ so that
\begin{equation}
{\bar h} \sim w \sim  (q-q_c)^{1/3} t^{1/3}.
\end{equation}

Within the simple mean-field approximation we were able
to observe a roughening transition occurring at the
transition line shown in Fig. \ref{phase}. Nevertheless,
this approximation is not able to predict a moving
surface with a constant velocity. This deficiency
will be overcome in the next section 
by means of the pair mean-field approximation.

\section{Pair mean-field approximation}

In the pair mean-field approach we solve the equation (\ref{two})
by using the approximation for the three-site probability distribution
\begin{equation}
P(h_1,h_2,h_3,t)= 
\frac{P(h_1,h_2,t)P(h_2,h_3,t)}{P(h_2,t)},
\end{equation}  
where $P(h_1,h_2,t)$ and $P(h_2,t)$ are the two-site and one-site
probability distribution. These two quantities are related by
\begin{equation}
P(h_2,t)=\sum_{h_1}P(h_1,h_2,t)=\sum_{h_3}P(h_2,h_3,t).
\end{equation}    

Due to the RSOS condition there are actually three
types of two-site probabilities: 
$P_{k,k}(t)$, 
$P_{k,k-1}(t)$ and 
$P_{k,k+1}(t)$,
where as before we are using the integer variable $k$ in the place of
the height $h$. Together with
the one-site probability $P_k$ they form a set of four variables.
However they are not all independent since $P_{k+1,k}=P_{k,k+1}$ 
and 
\begin{equation}
P_k= P_{k,k}+P_{k,k-1}+P_{k,k+1}.
\end{equation}
Therefore we are left with two independent variable, for
each $k$, which we choose to be $P_{k,k}$ and $P_{k,k+1}$.
Initially we have $P_{k,k+1}= 0$ for all $k$ and
$P_{k,k}=0$ for $k\neq 0$ and $P_{0,0}=1$.

\begin{figure}
\smallskip
\centering
\epsfig{file=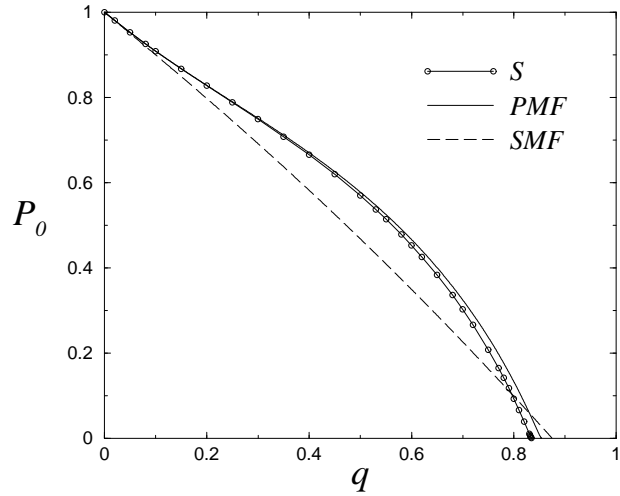,width=8cm}
\caption{Order parameter $P_0$ as a function of $q$ for $p=0.5$ obtained 
from simple mean-field (SMF), pair mean-field (PMF) approximation
and numerical simulations (S) performed with 2048 sites and
100 independent realizations.
The critical values are $q_c=0.875$ for SMF, $q_c=0.8539$ 
for PMF and $q_c=0.8346(1)$ for the simulations.}
\label{fop}	
\end{figure}

To simplify notation,
we denote $P_{k,k+1}$ by $y_k$ and $P_{k,k}$ by $x_k$ so that
\begin{equation}
P_k = x_k + y_k + y_{k-1}.
\label{red}
\end{equation}
The pair mean-field equations for $x_k$ and $y_k$,
corresponding to the model defined by (\ref{sm}), then reads
\[
\frac{d}{dt}x_k
= 2q\biggl[\frac{y_{k-1}(y_{k-1}+x_{k-1})}{P_{k-1}}
-\frac{x_k(x_k+y_k)}{P_k}\biggr]
\]
\begin{equation}
+2r\biggl[\frac{y_k(y_k+x_{k+1})}{P_{k+1}}
-\frac{x_ky_{k-1}}{P_k}\biggr]-2p\frac{x_k^2}{P_k}\gamma_k,
\label{xk}
\end{equation}
where we should set $x_{-1}=y_{-1}=0$ when $k=0$,
and
\begin{equation}
\frac{d}{dt}y_k= q \frac{x_k^2-y_k^2}{P_k}
- r\,\frac{y_k^2}{P_{k+1}}
+ p\,\frac{x_{k+1}^2}{P_{k+1}}.
\label{yk}
\end{equation}
The numerical integration of the coupled equations (\ref{xk}) and 
(\ref{yk}), performed by repeated iteration from the initial condition
$x_k= 0$ for $k\neq 0$, $x_0 = 1$ 
and $y_k= 0$ for all $k$,
allows us to determine the one-site probability distribution
(\ref{red}). From the obtained distribution we get 
the mean height and the interface width by the use
of Eqs. (\ref{observ}) and the interface velocity
by derivating numerically the mean height with respect to time.

The transition line is shown in Fig. \ref{phase}.
For comparison we also show the transition line
obtained from numerical simulation.
Below the transition line, in the pinned phase, the density of
sites $P_0$ in contact with the substrate is found to behave as 
\begin{equation}
P_0 \sim (q_c-q),
\end{equation}
so that the exponent $\beta=1$, the same value found
in the one-site mean-field approximation.
The width of the interface diverges at the critical line as
\begin{equation}
w \sim (q_c-q)^{-1/3},
\label{umtres}
\end{equation}
yielding an exponent $\zeta=1/3$.

\begin{figure}
\centering
\epsfig{file=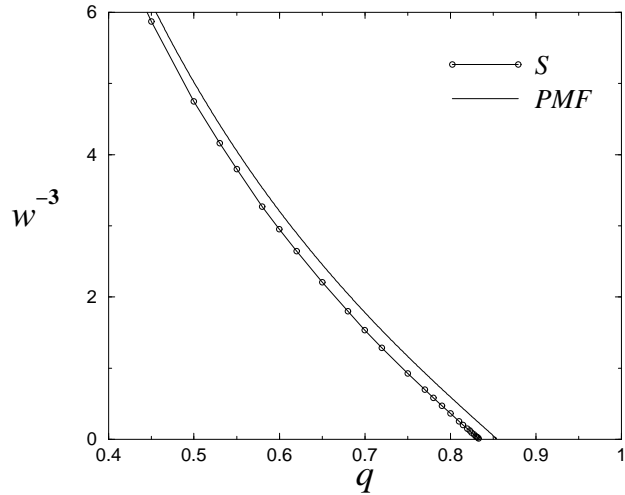,width=8cm}
\caption{Interface width $w$ as a function of $q$ for
$p=0.5$ obtained from pair mean-field (PMF) approximation
and numerical simulations (S) performed with 2048 sites
and 100 independent realization.
The critical values are $q_c=0.8539$ and $q_c=0.8346(1)$, 
respectively.}
\label{fw}
\end{figure}

As an example, we show in Figs. \ref{fop} and \ref{fw},
respectively, the order parameter $P_0$ and the width $w$
as functions of the deposition rate $q$, at $p=0.5$.
We see that $P_0$ is linear near the transition point 
and that $w^{-3}$ is linear supporting the behavior (\ref{umtres}).

\begin{figure}
\centering
\epsfig{file=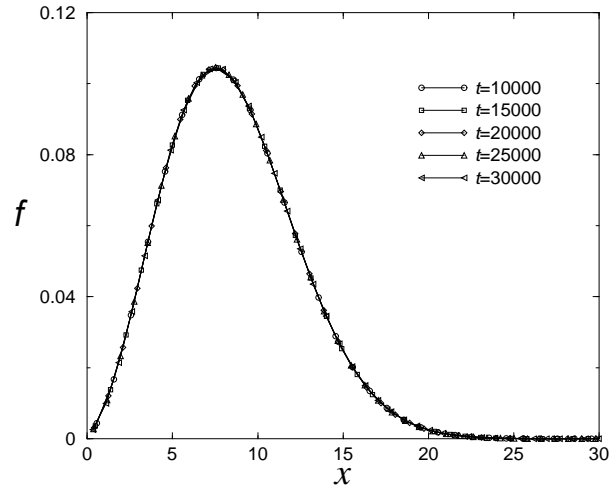,width=8cm}
\caption{Scaling function $f= P(h,t)\,t^{1/4}$ versus
$x=h\,t^{-1/4}$ as obtained from the pair
mean-field approximation for the values of $t$
shown in figure, at the critical point $q=p=1$.
The function $f(x)$ extrapolates to zero when $x\to 0$.}
\label{few}
\end{figure}

\begin{figure}
\centering
\epsfig{file=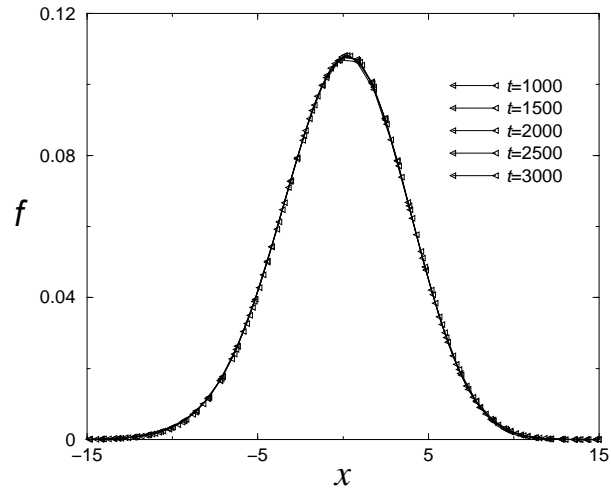,width=8cm}
\caption{
Scaling function $f= P(h,t)\, t^{1/3}$ versus
$x=(h-vt)\,t^{-1/3}$ as obtained from the pair
mean-field approximation for the values of $t$
shown in figure, inside the moving phase at the point $q=2$ and $p=1$.
The velocity is found to be $v=0.112$.}
\label{fkpz}
\end{figure}

Above and at the transition line, 
the numerical integration gives 
a time dependent probability distribution $P_k(t)$
from which we calculate the average height and the
width of the interface. 
Along the transition line we found that
\begin{equation}
w \sim t^{1/4}, \qquad\qquad q=q_c,
\label{50}
\end{equation}
and above it
\begin{equation}
w \sim t^{1/3}, \qquad\qquad q>q_c.
\label{51}
\end{equation}
In Figs. \ref{few} and \ref{fkpz}
we show a data collapse of $P_k(t)$
by using the scaling form (\ref{scal}) 
at the critical point ($p=1$ and $q=1$) and above the critical point,
inside the moving phase ($p=1$ and $q=2$).
In the former case we have used $\gamma=1/4$ 
and in the latter, $\gamma=1/3$.
The good data collapse leads us to the results
(\ref{50}) and (\ref{51}).

The asymptotic results for the width of the interface
are the same as those found by means of the one-site
mean-field approximation. However, as already mentioned,
the two-site approximation is capable
of describing a moving interface.
We found that, inside the moving phase the mean height increases
with a constant velocity $v$ so that the pair mean-field
approximations predicts correctly a linear growth of
the moving interface.
As we approach the critical line the velocity vanishes
according to
\begin{equation}
v \sim (q-q_c),
\end{equation}  
that is, $\theta=1$.

As we have seen, the pair mean-field approximation takes into account
the correlation of two consecutive sites which results in 
the exact satisfaction of the RSOS condition
in opposition to the simple mean-field approximation. 
Because of that the pair mean-field results are
closer to the simulation ones when compared to the
simple mean-field results. For instance, the transition line
is in general closer to the simulation results 
and has the right concavity. 

\section{Detailed balance}

When detailed balance is obeyed, it turns out that the 
two-site approximation becomes an exact solution.
In this case the stationary probability distribution is of the 
form (\ref{pt}) and  the elements of $X_k$  and $Y_k$ of the 
transfer matrix $T$ are related to $x_k$ and $y_k$ by
$x_k=X_k P_k/\Lambda$ and $y_k=Y_k\sqrt{P_kP_{k+1}}/\Lambda$,
where $\Lambda$ is the dominant eigenvalue of $T$.
The detailed balance conditions (\ref{db1}), (\ref{db2}) and (\ref{db3}) 
are then equivalent to the following relations: 
\begin{equation}
p=r=1,
\qquad\qquad
q \frac{x_k}{P_k} = \frac{x_{k+1}}{P_{k+1}},
\qquad\qquad
y_k^2 = x_k x_{k+1},
\end{equation}
from which follows the solution
\begin{equation}
\frac{x_k}{P_k} = \frac{1}{\lambda} q^k,
\qquad\qquad
\frac{y_k}{\sqrt{P_k P_{k+1}}} = \frac{1}{\lambda} q^{k+1/2},
\end{equation}
where $\lambda$ is a constant to be found.
That this is indeed a solution of the
pair mean-field equations can be checked by substitution.

Inserting these relations into equation (\ref{red}) we get the
following eigenvalue equation for $\phi_k = \sqrt{P_k}$
\begin{equation}
q^{k-1/2}\, \phi_{k-1}+q^k\, \phi_k +q^{k+1/2}\, \phi_{k+1}
= \lambda\, \phi_k
\label{eigen}
\end{equation}
so that the constant $\lambda$ is identified as the
eigenvalue, actually, the dominant eigenvalue $\Lambda$.
The dominant eigenfunction gives the one-site probability
distribution $P_k=\phi_k^{\,2}$. The eigenvalue equation (\ref{eigen})
was found by Hinrichsen et al. \cite{hinr03} who solved it in the
vicinity of the critical point by using a continuous height
approximation. Their solution gives the following results
in the vicinity of the critical point $q=q_c=1$
\begin{equation}
w \sim (q_c-q)^{-1/3},
\end{equation}
and 
\begin{equation}
P(0) \sim (q_c-q).
\end{equation}
These results are the same results found in the previous section
by numerical integration of the two-site pair approximation
equations.

\section{Conclusion}

We have studied a lattice model for surface growth
by means of one-site and two-site mean-field approximations.
Within the simple mean-field approximation,
whose equations were solved exactly, 
we were able to observe a roughening transition occurring
with growth exponents equal to 1/4 at the transition line
and 1/3 inside the rough phase.
This simple approximation was not able to predict a moving
surface with a constant velocity. However, this deficiency
was overcome by means of the pair mean-field approximation
which predicts correctly a linear growth of the moving interface.
In general, the pair approximation
gives results that are in better accordance with 
numerical simulations. 
The critical line is closer to that
given by numerical simulations. The growth exponent is 1/4 
at the critical line and 1/3 inside the moving phase.
At $p=1$ this approximations correctly predicts the crossover
from the EW, at $q=1$, to KPZ, for $q > 1$. 

The pair mean-field approximations gives also the following results.
As one approaches the critical line from inside the pinned
phase, the order parameter vanishes with an exponent $\beta=1$
and the width of the interface diverges with an
exponent $\zeta=1/3$. If one approaches the line
from the moving phase, the velocity of the interface
vanishes with an exponent $\theta=1$.

The pair mean-field approximation has two important features:
it takes into account the RSOS condition exactly
and reduces to the exact solution when detailed balance is fulfilled.
This last feature indicates that the pair mean-field
is a good approximation for the nonequilibrium case 
($p\neq 1$), where exact results are generally not known.
This approximation can also be use to get the properties
and the phase diagram of the general model defined in
Fig. \ref{rate}, with or without a wall.
For growth models that do not respect the RSOS condition, 
in principle, the pair approximation can also be used but
the equations will have a more cumbersome form.


\section*{Acknowledgment}

This research was supported by the Brazilian agency CNPq.


\end{document}